\documentclass[traditabstract]{aa}

\usepackage{graphicx}
\usepackage{txfonts}
\usepackage{natbib}
  \bibpunct{(}{)}{;}{a}{}{,}
\usepackage{lscape}

\newcommand{\feh}{\mathrm{[Fe/H]}}
\newcommand{\teff}{T_\mathrm{eff}}
\newcommand{\logg}{\log g}

\newcommand{\tc}{T_\mathrm{C}}

\begin{document}

\defcitealias{melendez09:twins}{M09}
\defcitealias{ramirez09}{R09}
\defcitealias{reddy03}{R03}
\defcitealias{allende04:s4n}{A04}
\defcitealias{takeda07:abundances}{T07}
\defcitealias{neves09}{N09}
\defcitealias{gonzalez10}{G10}
\defcitealias{bensby11}{BFO}

\title{A possible signature of terrestrial planet formation \\ 
in the chemical composition of solar analogs}

\titlerunning{A signature of terrestrial planet formation?}
\authorrunning{Ram\'{\i}rez et al.}

\author{I. Ram\'{\i}rez   \inst{1} \and
	M. Asplund        \inst{1} \and
	P. Baumann        \inst{1} \and
	J. Mel\'endez     \inst{2} \and
	T. Bensby         \inst{3}
       }

\institute{Max Planck Institute for Astrophysics,
           Postfach 1317, 85741 Garching, Germany \\
	   \email{ivan,asplund,pbaumann@mpa-garching.mpg.de}
	   \and
	   Centro de Astrof\'{\i}sica da Universidade do Porto, 
	   Rua das Estrelas 4150-762 Porto, Portugal \\
	   \email{jorge@astro.up.pt}
	   \and
	   European Southern Observatory,
	   Alonso de Cordova 3107, Vitacura, Casilla 19001, 
	   Santiago 19, Chile \\
	   \email{tbensby@eso.org}
          }

\date{Received 18/03/2010; accepted 13/08/2010}

\abstract
{
Recent studies have shown that the elemental abundances in the Sun are anomalous when compared to most (about 85\,\%) nearby solar twin stars. Compared to its twins, the Sun exhibits a deficiency of refractory elements (those with condensation temperatures $\tc\gtrsim900$\,K) relative to volatiles ($\tc\lesssim900$\,K). This finding is speculated to be a signature of the planet formation that occurred more efficiently around the Sun compared with the majority of solar twins. Furthermore, within this scenario, it seems more likely that the abundance patterns found are specifically related to the formation of terrestrial planets. In this work we analyze abundance results from six large independent stellar abundance surveys to determine whether they confirm or reject this observational finding. We show that the elemental abundances derived for solar analogs in these six studies are consistent with the $\tc$ trend suggested as a planet formation signature. The same conclusion is reached when those results are averaged heterogeneously. We also investigate the dependency of the abundances with first ionization potential (FIP), which correlates well with $\tc$. A trend with FIP would suggest a different origin for the abundance patterns found, but we show that the correlation with $\tc$ is statistically more significant. We encourage similar investigations of metal-rich solar analogs and late F-type dwarf stars, for which the hypothesis of a planet formation signature in the elemental abundances makes very specific predictions. Finally, we examine a recent paper that claims that the abundance patterns of two stars hosting super-Earth like planets contradict the planet formation signature hypothesis. Instead, we find that the chemical compositions of these two stars are fully compatible with our hypothesis.
}

\keywords{stars: abundances --
          Sun: abundances --
	  stars: planetary systems
	 }

\maketitle

\section{Introduction}

Finding planets outside the solar system is one of the major endeavors of contemporary astrophysics. Large efforts to detect them using radial velocity, transits, and microlensing observations are currently underway as they have been proven successful. Indeed, more than 450 extrasolar planets have been discovered to date. A majority of the planets that have been detected have masses similar to that of Jupiter and a significant fraction of them are in close-in, short-period orbits, hence their classification as ``hot Jupiters'' \citep[see, e.g., the review by][]{udry07}. The parent stars are mostly G and K dwarfs. Due to the intrinsic limitations of the techniques currently used to find exoplanets, however, terrestrial planets in Earth-like orbits have remained elusive. This situation may change in the next few years as transiting data from the \textit{Kepler} mission \citep{basri05} are scrutinized spectroscopically. In the meantime, we face the challenge of developing new practical techniques to find those objects.

Comparisons of the chemical composition of stars that host extrasolar planets and stars for which planets have not yet been detected have been performed by a number of groups \citep[e.g.,][]{gonzalez01,sadakane02,heiter03,santos04,fischer05}. These studies unanimously agree on the metal-rich nature of planet hosts (i.e., stars known to host planets are on average more metal-rich than a typical field star), first suggested by \cite{gonzalez97}. From a practical point of view, this result can be used to increase the efficiency of exoplanet searches. Note, however, that it is not clear whether evolved stars which host planets are, on average, more metal-rich than a typical field star \citep[e.g.,][]{hekker07,pasquini07,takeda08}.

Attempts have been made to find other indicators in the photospheric chemical composition of stars that would suggest the presence of extrasolar planets. For instance, it has been suggested that the lithium abundances in stars that host planets are more depleted than those in single stars \citep[e.g.,][]{chen06,gonzalez08,israelian09} as a result of a star-planet interaction that has led to enhanced mixing and/or deepening of the convective zone. This, in turn, increases the rate of lithium burning in the stellar interior. It is possible, however, that these results have been affected by a combination of heterogeneous abundance results or, more importantly, observational biases related to stellar age and metallicity \citep{ryan00,luck06,melendez09:lithium,baumann10}. Therefore, the possible connection between lithium abundance and presence of planets is not conclusive.

\citet[hereafter \citetalias{melendez09:twins}]{melendez09:twins} have shown that the chemical composition of the Sun is anomalous when compared to stars of very similar fundamental parameters, so called ``solar twins.'' Inspection of abundance trends with condensation temperature ($\tc$) revealed, for the first time, that, compared to most solar twins (about 85\,\%), the Sun has a deficiency of refractory elements ($\tc\gtrsim900\,$K) relative to volatiles ($\tc\lesssim900\,$K). In the solar photosphere, elements with the highest $\tc$ are depleted by about 20\,\% relative to volatiles. A 20\,\% difference in relative abundances (about 0.08\,dex in the standard [X/H] scale)\footnote{We use the standard notation: $A_\mathrm{X}=\log n_\mathrm{
X}/n_\mathrm{H}$+12, where $n_\mathrm{X}$ is the number density of the element X
; $\mathrm{[X/H]}=A_\mathrm{X}-A_\mathrm{X}^\odot$; and $\mathrm{[X/Fe]}=\mathrm
{[X/H]}-\mathrm{[Fe/H]}$.} is extremely difficult to detect because systematic errors are often of similar magnitude or larger. In the \citetalias{melendez09:twins} study, those errors were minimized by analyzing only stars with fundamental parameters very similar to solar, for which systematic errors in the derived abundances cancel-out in a strictly differential analysis. Using a different sample of stars and a new set of observations (solar twins in the northern hemisphere as opposed to southern hemisphere stars as in the \citetalias{melendez09:twins} case), \citet[hereafter \citetalias{ramirez09}]{ramirez09} determined abundance trends with condensation temperature which are in excellent agreement with those found by \citetalias{melendez09:twins}. Both the amplitude of the effect ($\simeq20$\,\% deficiency of refractories relative to volatiles) and the frequency of stars in which the trend is observed ($\simeq85$\,\%) were confirmed by this latter study. More recently, \cite{gonzalez10} have confirmed both \citetalias{melendez09:twins} and \citetalias{ramirez09} results using a different sample of stars and approach to the analysis of the stellar abundance data. In addition, they explore the impact of effective temperature and metallicity on these abundance trends.

\citetalias{melendez09:twins} have speculated that this ``solar abundance anomaly'' is due to the formation of planets in the solar system, and possibly specifically the terrestrial planets,  because the missing refractories are currently inside them. If this hypothesis is proven correct, very precise relative abundances could be used to identify stars that harbor terrestrial planets. A lack of correlation in the abundances relative to solar with $\tc$ would indicate the presence of such systems around other stars. The amount of observing time required to identify planet hosts with this ``abundance method'' is orders of magnitude smaller than that necessary to detect planets using radial velocity or transit techniques. Only one set of observations is required per star, in contrast to the lengthy campaigns that have to be carefully designed to detect very small radial velocity or photometric variations.

The identification of terrestrial planet hosts using abundances will help us to answer important questions such as (1) how rare is the solar system? and (2) which types of stars are more likely to host terrestrial planets? However, the connection between the abundance trends with condensation temperature and terrestrial planet formation needs to be explored further.

A large number of stellar photospheric chemical composition studies have been published to date. Only a few, however, are based on a homogeneous analysis of high quality data for large samples of stars and deal with more than a handful of chemical elements. In this paper, we compile results from six of those studies. We concentrate on stars that are similar to the Sun, since the signature is difficult to detect and possibly blurred by systematic errors in larger samples including stars with significantly different fundamental parameters. We analyze the trends with condensation temperature for these samples and check whether other independent studies of stellar abundances support the observational findings of \citetalias{melendez09:twins} and \citetalias{ramirez09}.

\section{Sample selection and abundances adopted} \label{s:samples}

We have compiled abundance results from six independent studies: \citet[hereafter \citetalias{reddy03}]{reddy03}, \citet[hereafter \citetalias{allende04:s4n}]{allende04:s4n}, \citet[hereafter \citetalias{takeda07:abundances}]{takeda07:abundances}, \citet[hereafter \citetalias{neves09}]{neves09}, \citet[hereafter \citetalias{gonzalez10}]{gonzalez10}, and \citeauthor{bensby11} (in preparation; hereafter \citetalias{bensby11}). A common characteristic among these studies is the homogeneity of the analysis and the high quality of the data employed. Each has adopted a unique temperature scale, spectral line selection, and method of abundance determination. In addition, abundances were determined from equivalent width matching and/or line-profile fitting using spectra of high resolution ($R\gtrsim50,000$) and high signal-to-noise ratio ($R\gtrsim100$). In all cases, abundances for at least 10 elements were determined.

\begin{table}
\centering
\caption{Samples analyzed in this paper, selection criteria for solar analogs, and number of stars selected.}
\begin{tabular}{lcc}\hline\hline
Sample & Solar analogs range & Number \\ 
       & $\Delta\teff/\Delta\logg/\Delta\feh$ & of stars \\ \hline
\cite{reddy03}             & $\pm300$\,K / $\pm0.2$ / $\pm0.2$ & 11 \\
\cite{allende04:s4n}       & $\pm150$\,K / $\pm0.2$ / $\pm0.2$ & 12 \\
\cite{takeda07:abundances} & $\pm150$\,K / $\pm0.2$ / $\pm0.2$ & 14 \\
\cite{neves09}             & $\pm100$\,K / $\pm0.1$ / $\pm0.1$ & 19 \\
\cite{gonzalez10}          & $\pm100$\,K / $\pm0.1$ / $\pm0.1$ & 11 \\
\citeauthor{bensby11} (in preparation) & $\pm100$\,K / $\pm0.1$ / $\pm0.1$ & 20 \\ \hline
\end{tabular}
\label{t:samples}
\end{table}

\begin{figure*}
\includegraphics[width=15.5cm,bb=70 365 760 770]
{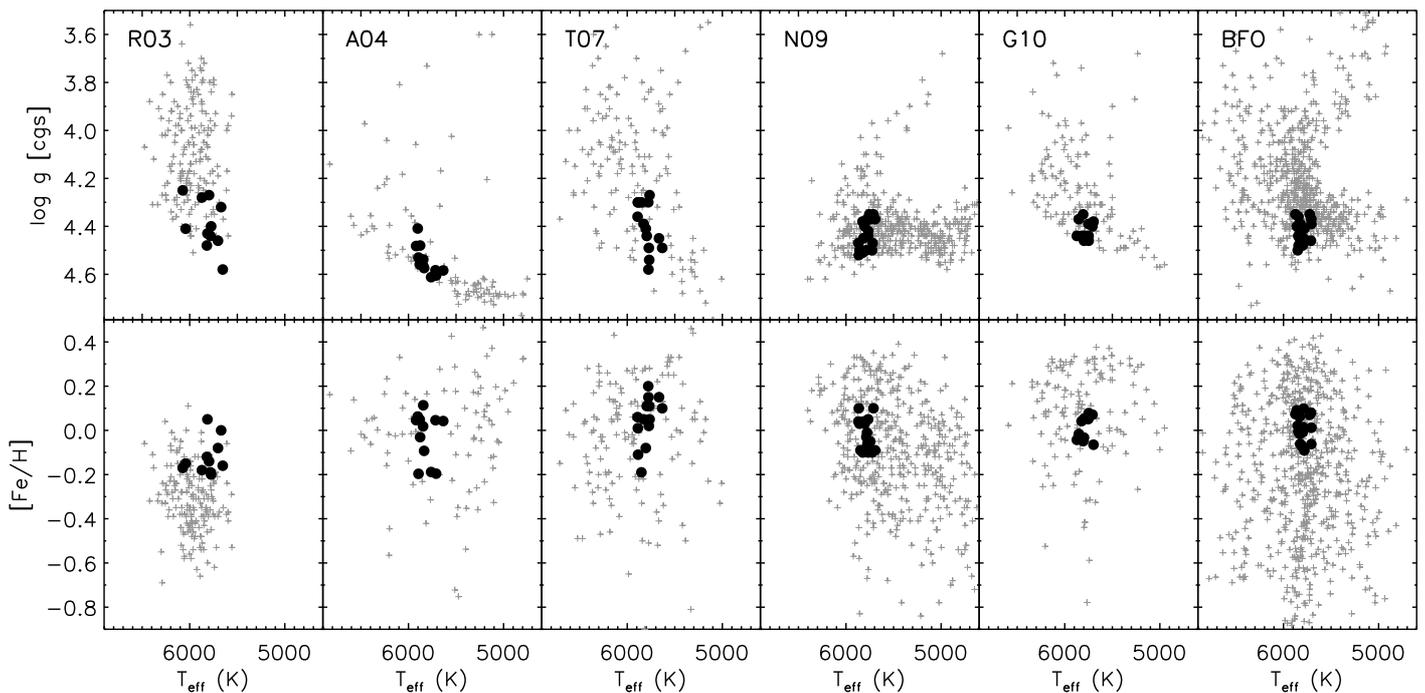}
\caption{Upper panels: effective temperature ($\teff$) versus surface gravity ($\logg$) for all stars in the  \citetalias{reddy03}, \citetalias{allende04:s4n}, \citetalias{neves09}, \citetalias{gonzalez10}, and \citetalias{bensby11} studies (crosses). Filled circles represent the solar analogs used in this work. Lower panels: $\teff$ versus [Fe/H] for the stars shown in the upper panels.}
\label{f:samples}
\end{figure*}

From each sample, only those stars with fundamental parameters ($\teff,\logg,\feh$) within a certain range from the solar values ($\teff=5777\,\mathrm{K},\logg=4.44,\feh=0$) were selected. In the \citetalias{ramirez09} study, the ranges adopted were $\pm100$\,K in effective temperature, $\pm0.1$\,dex in surface gravity, and $\pm0.1$\,dex in iron abundance. An identical selection was made for the \citetalias{neves09}, \citetalias{gonzalez10}, and \citetalias{bensby11} samples. For the \citetalias{reddy03}, \citetalias{allende04:s4n}, and \citetalias{takeda07:abundances} samples, however, we used wider ranges of allowed stellar parameters in order to have at least 10 solar analogs, as detailed in Table~\ref{t:samples}. The number of stars selected from each study are listed also in Table~\ref{t:samples}. In Fig.~\ref{f:samples} we illustrate the overall properties of each sample, highlighting the location of the stars selected for this work.

\citetalias{reddy03} used spectra obtained with the Robert~G.~Tull spectrograph on the 2.7\,m Telescope at McDonald Observatory. The main goal of this work was to study Galactic chemical evolution using 181 nearby stars, almost all of them members of the Galactic thin disk. They derived abundances for 27 elements. The \citetalias{reddy03} results were complemented with data for 176 stars with high probability of being thick-disk members by \cite{reddy06}. If we include the data from this later study in our work, the precision of the average abundance ratios that we derive for solar analogs (cf.~Sect.~\ref{s:tc_trend}) decreases from about 0.05 to 0.09\,dex. The abundance trend we find with condensation temperature becomes more noisy but its shape does not change significantly. We did not include the \cite{reddy06} results in our work because we require the highest precision possible. Small changes in the determination of stellar parameters between \citetalias{reddy03} and \cite{reddy06} and the slightly lower quality of the data employed by the latter may be responsible for the increased scatter of the combined sample.

\citetalias{allende04:s4n} performed a volume limited survey of 118 stars more luminous than $M_V=6.5$\,mag within a 15\,pc radius from the Sun. Their data were obtained at McDonald Observatory and with the FEROS spectrograph on the 1.52\,m Telescope at La Silla. \citetalias{allende04:s4n} derived abundances for 16 elements. The spectra from the \citetalias{allende04:s4n} survey are available online.\footnote{Spectroscopic Survey of Stars in the Solar Neighborhood (S$^4$N): http://leda.as.utexas.edu/s4n/.} We re-determined the stellar parameters $\teff,\logg,\feh$ of a number of solar analogs using these data and the methods described in \citetalias{ramirez09}. We found a mean difference of 152\,K in $\teff$ and 0.06\,dex in $\logg$ between the re-derived parameters and those given by \citetalias{allende04:s4n}, with the latter being systematically smaller. Indeed, the zero point of the temperature scale adopted by \citetalias{allende04:s4n} has been recently revised and found to be significantly cooler than previously thought \citep{casagrande10}. \citetalias{allende04:s4n} provide detailed tables that allowed us to compute corrections to the abundances of all elements given an effective temperature correction. Thus, we applied a +152\,K correction to the parameters and abundances from \citetalias{allende04:s4n}. We did not correct the abundances for the shift in $\logg$ values because there are no corresponding correction tables. The selection of solar analogs from this study was made after the $\teff$ correction to the abundances was applied. This type of correction was not possible for the other samples because the observational data are not publicly available, thus preventing us from determining the temperature correction required, and because none of the other studies provide detailed tables of abundance corrections for given shifts in fundamental parameters. We discuss the implication of this correction for the \citetalias{allende04:s4n} results to our work in Sect.~\ref{s:tc_trend}.

Fundamental parameters and elemental abundances of 15 elements for 160 FGK dwarf and subgiant stars observed with the 1.88\,m Telescope at Okayama Astrophysical Observatory were presented by \citetalias{takeda07:abundances}. This work discusses chemical evolution, comparison of abundances in planet-host and single stars, and empirical tests of the validity of the LTE approximation in abundance studies.

Using spectra from the HARPS/GTO planet search programs, \citetalias{neves09} determined abundances of 12 elements in 451 stars. The stellar parameters $\teff,\logg,\feh$ adopted in this study are from \cite{sousa08}. The wavelength coverage and typical S/N of the HARPS data is unfortunately not appropriate to derive precise abundances of the volatile elements C, N, and O. However, very precise abundances for many refractory elements are given by \citetalias{neves09}.

For more than a decade, G.\ Gonzalez and collaborators have collected high quality spectra and analyzed them homogeneously to derive precise stellar parameters and abundances with the goal of investigating the connection between stellar abundances and extrasolar planets \citep[e.g.,][]{gonzalez97,gonzalez98,gonzalez01}. In their most recent publication \citepalias{gonzalez10}, the abundance correlation with condensation temperature is explored for about 85\,\% of their 85 stars with planets and 59 stars not known to host planets. The approach used in \citetalias{gonzalez10} consists of a differential comparison of abundance trends weighted by the distance between stars in stellar parameter space, as defined in \cite{gonzalez08}. In this way they are able to perform differential analyses not restricted to solar twins. Abundances for ten elements are available from \citetalias{gonzalez10} although a comment must be made about their Sc abundances, which are based on two \ion{Sc}{ii} lines. In \citetalias{ramirez09}, three \ion{Sc}{i} lines were preferred. The reason was that the star-to-star scatter obtained by \citetalias{ramirez09} for their solar twins reduced from 0.038\,dex for \ion{Sc}{ii} to 0.018\,dex for \ion{Sc}{i}. The average [Sc/Fe] abundance ratio for \ion{Sc}{i} lines was 0.024\,dex while that for \ion{Sc}{ii} lines was $-0.003$. Although non-LTE effects are expected to be stronger for the \ion{Sc}{i} lines, the [\ion{Sc}{i}/Fe] ratio, which is the quantity that we use in this paper, might be less sensitive to non-LTE because the iron abundances are dominated by the large number of \ion{Fe}{i} lines. It is likely that the systematic errors are more similar between \ion{Fe}{i} and \ion{Sc}{i} lines than between \ion{Fe}{i} and \ion{Sc}{ii}, making \ion{Sc}{i} lines more reliable for strictly differential analysis. We proceed our study including the Sc abundances from \citetalias{gonzalez10} but we will also comment on how their exclusion, justified by the statements made above, affects the conclusions.

\cite{bensby03,bensby05} presented abundances for 12 elements in 102 F and G dwarf stars in the solar neighborhood based on high-resolution spectra obtained with the FEROS/ESO 1.5m, SOFIN/NOT, and UVES/VLT spectrographs. Expanding those studies, the \citetalias{bensby11} study will contain a total of 703 F and G dwarf stars, adding $\simeq600$ new stars that have been observed with the MIKE spectrograph on the 6.5\,m Clay Magellan telescope. These new spectra cover the full optical wavelength regime (350-920\,nm), have high spectral resolution ($R\simeq65,000$), and high signal-to-noise ratio ($S/N>250$). As the analysis method in \citetalias{bensby11} is very similar to \cite{bensby03,bensby05}, the uncertainties in stellar parameters and abundance ratios are similar as well. First results from this expanded data set have been presented in \cite{bensby07:hercules,bensby07}, \cite{feltzing08}, and \cite{bensby10:rio}. The relevant atomic data adopted by \citetalias{bensby11} are given in \cite{bensby03}.

The oxygen abundances from the \citetalias{bensby11} work were derived by them from the 777\,nm O~\textsc{i} triplet lines, which are known to be severely affected by non-LTE \citep[e.g.,][]{fabbian09}. Although the range in stellar parameters adopted for the solar analogs in this sample is small, \citetalias{ramirez09} have found that differential non-LTE corrections slightly improve the results. Therefore, we corrected the oxygen abundances from \citetalias{bensby11} using the non-LTE correction tables by \cite{ramirez07}. We find that the star-to-star scatter in the average [O/Fe] abundance ratio that we derive for the solar analogs from the \citetalias{bensby11} study decreases from 0.048\,dex in LTE to 0.038\,dex in non-LTE.  We did not use the more recent non-LTE corrections by \cite{fabbian09} because their tables do not cover $\feh>0$. The differential non-LTE correction (star relative to Sun) by \cite{ramirez07} for a star that is 100\,K warmer than the Sun (the other parameters being solar) is about 0.01\,dex greater than that by \cite{fabbian09}. A similar calculation for a $\feh=-0.1$ star yields a difference of --0.02\,dex. The impact of $\logg$ differences is negligible within the range allowed for solar analogs in this paper. These differences are small but could have an impact on the average [O/Fe] abundance ratios we derive. Unfortunately, we are unable to use \cite{fabbian09} non-LTE correction tables for the reason exposed above.

As shown by \cite{bensby03,bensby05}, abundance ratios of kinematically selected thin- and thick-disk stars are very similar at $\feh\simeq0.0$. This is not the case at lower metallicities. We checked that the results presented in Sect.~\ref{s:tc_trend} are not affected by the inclusion of thick-disk solar analogs (i.e., excluding them did not have a significant impact). Only about 15\,\% of the stars studied in this paper have high kinematic probability of being thick-disk members, a number that is roughly consistent with current estimates of the fraction of thick-disk stars in the solar neighborhood.

\section{Abundances versus condensation temperature} \label{s:tc_trend}

\begin{figure}
\includegraphics[width=9.2cm,bb=50 350 480 1375]
{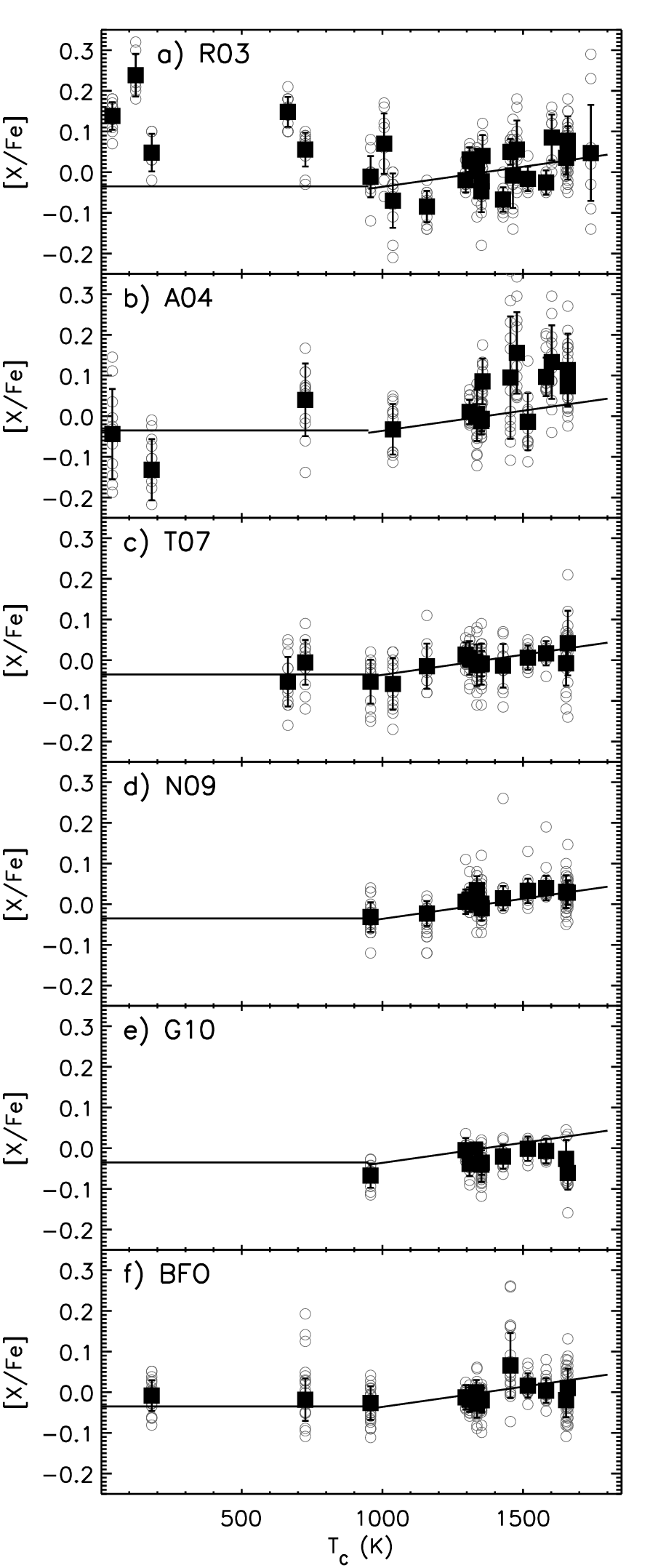}
\caption{Abundance ratios relative to iron for the solar twins and analogs from the six studies used in this paper as a function of condensation temperature (open circles). Filled squares with error bars represent the average value and standard deviation for each element. Solid lines represent the average [X/Fe] value of volatiles ($\tc<900$\,K) and the linear fit to the [X/Fe] versus $\tc$ relation of refractories ($\tc>900$\,K) derived for solar twins by \cite{ramirez09}.}
\label{f:abund_tcond}
\end{figure}

For each of the six samples analyzed in this paper we determined an average [X/Fe] vs.\ $\tc$ relation. For each element X, we calculated an average [X/Fe] and standard deviation with the data for the solar analogs (weighted by the observational errors, when available). We adopted the 50\,\% equilibrium condensation temperatures for a solar system composition gas computed by \cite{lodders03}. The resulting average [X/Fe] vs.\ $\tc$ relations are shown in Fig.~\ref{f:abund_tcond}, along with the corresponding relation found by \citetalias{ramirez09}. Table~\ref{t:abund_tcond} lists the average [X/Fe] abundance ratios of solar analogs that we obtain for each of the six studies used in this paper. We also list in Table~\ref{t:abund_tcond} our adopted condensation temperature and first ionization potential for each element. The latter are adopted from the ``Atomic Properties of the Elements" compilation by the National Institute of Standards and Technology (NIST).\footnote{Available online at http://www.nist.gov/physlab/data/periodic.cfm} In Table~\ref{t:abund_tcond_rm09} we show the average [X/Fe] abundance ratios of solar twin stars from the \citetalias{melendez09:twins} and \citetalias{ramirez09} studies. The adopted condensation temperatures and first ionization potentials for the elements studied in these two papers are given in Table~\ref{t:abund_tcond}, except for element P (Z=15), which is only covered by the \citetalias{melendez09:twins} study ($\tc=1229$\,K, $\mathrm{FIP}=10.49$\,eV).

The abundances of volatile elements in the solar analogs from the \citetalias{reddy03} study are in severe disagreement with those found by \citetalias{ramirez09} but their $\tc$ trends for the refractory elements are in reasonably good agreement (Fig.~\ref{f:abund_tcond}a). In this study, the spectroscopically determined abundances, relative to iron, of carbon, nitrogen, and oxygen, which are the elements with the lowest $\tc$ in Fig.~\ref{f:abund_tcond}a, are very sensitive to errors in the $\teff$ scale adopted. The $\teff$ sensitivity of these elements is stronger than that of other elements. \citetalias{reddy03} noticed, for example, that their [C/Fe] abundance ratios are systematically higher by 0.15\,dex compared to those by \cite{edvardsson93}, and attributed the difference to an offset of about 80\,K in the $\teff$ scales adopted. Their [N/Fe] abundance ratios were determined from the analysis of two weak N lines. Errors in the continuum determination of weak lines may lead to severely overestimated equivalent widths when noisy data are used. In these cases, relative abundances can be systematically off if the reference spectrum (the solar one) has much higher signal-to-noise ratio, as is often the case. Indeed, \citetalias{reddy03} noted that their N abundances are about 0.2\,dex higher on average compared to those determined by \cite{clegg81}, who used also atomic N lines. The other two elements that depart significantly from the \citetalias{ramirez09} result are Zn and S. In addition to potential systematic and statistical errors which could have resulted in \citetalias{reddy03} abundances being significantly higher than those by \citetalias{ramirez09}, it should be noted that this sample is, on average, slightly more metal-poor than the Sun (Fig.~\ref{f:samples}). This behavior is not observed in any of the other samples of solar analogs, which are shown to distribute uniformly around $\feh=0$. Thus, it is also possible that small chemical evolution effects are partially responsible for the discrepancy observed. Indeed, the four lowest $\tc$ elements from the \citetalias{reddy03} work show decreasing abundance ratios with $\feh$, although the slopes are dissimilar for the different elements (see Fig.~9 in \citetalias{reddy03}).

The \citetalias{allende04:s4n} abundances seem to broadly agree with the solar abundance anomaly reported by \citetalias{melendez09:twins} and \citetalias{ramirez09}, despite the relatively large uncertainties (Fig.~\ref{f:abund_tcond}b). On average, volatiles have $\mathrm{[X/Fe]}\simeq-0.06$ whereas refractories have an average [X/Fe] of about 0.02 (weighted average). The $\tc$ trend appears to be steeper compared to \citetalias{ramirez09} but the overall behavior of the abundances with $\tc$ is qualitatively similar. Since we shifted the parameters and abundances from \citetalias{allende04:s4n} following a 152\,K offset in $\teff$ (Sect.\ \ref{s:samples}), it is important to investigate the results using the original values. In this case the average abundance ratio of volatiles is about 0.08\,dex whereas that corresponding to the refractories is about 0.09\,dex. The linear $\tc$ trend for the refractories is conserved but the difference in average abundance between volatiles and refractories decreases significantly. The result using the original abundances resembles somewhat the \citetalias{reddy03} result (although the abundances of volatiles are not that high in the \citetalias{allende04:s4n} case), which reinforces our suspicion that the abundances of volatiles in these two studies (the original ones in the \citetalias{allende04:s4n} case) are too high due to errors in the $\teff$ scale. As mentioned in Sect.\ \ref{s:samples}, the $\teff$ scale adopted in these two studies, namely that by \cite{alonso96}, has been recently shown to suffer from a  relatively large systematic error ($\simeq100$\,K) in its zero point \citep{casagrande10}.

\begin{table*}
\caption{Average (star-to-star) elemental abundance ratios [X/Fe] of solar analogs in each of the samples studied in this work. Error bars are the weighted 1-$\sigma$ scatter.}
\centering
\begin{tabular}{rlrrrrrrrr}\hline\hline
Z & El. & $\tc$ & FIP & \multicolumn{1}{c}{\citetalias{reddy03}} & 
\multicolumn{1}{c}{\citetalias{allende04:s4n}} &
\multicolumn{1}{c}{\citetalias{takeda07:abundances}} & \multicolumn{1}{c}{\citetalias{neves09}} & \multicolumn{1}{c}{\citetalias{gonzalez10}} & \multicolumn{1}{c}{\citetalias{bensby11}} \\ 
 & & (K) & (eV) & & & & & & \\ \hline 
 6 &  C &   40 &  11.26 & $ 0.138\pm0.034$ & $-0.044\pm0.111$ &          $\dots$ &          $\dots$ &          $\dots$ &          $\dots$ \\
 7 &  N &  123 &  14.53 & $ 0.238\pm0.052$ &          $\dots$ &          $\dots$ &          $\dots$ &          $\dots$ &          $\dots$ \\
 8 &  O &  180 &  13.62 & $ 0.048\pm0.046$ & $-0.132\pm0.075$ &          $\dots$ &          $\dots$ &          $\dots$ & $-0.008\pm0.038$ \\
11 & Na &  958 &   5.14 & $-0.011\pm0.051$ &          $\dots$ & $-0.053\pm0.054$ & $-0.032\pm0.037$ & $-0.067\pm0.030$ & $-0.027\pm0.041$ \\
12 & Mg & 1336 &   7.65 & $ 0.005\pm0.052$ & $ 0.002\pm0.063$ & $-0.012\pm0.051$ & $ 0.035\pm0.034$ &          $\dots$ & $-0.028\pm0.035$ \\
13 & Al & 1653 &   5.99 & $ 0.035\pm0.041$ &          $\dots$ & $-0.008\pm0.055$ & $ 0.031\pm0.040$ & $-0.026\pm0.046$ & $-0.020\pm0.042$ \\
14 & Si & 1310 &   8.15 & $ 0.031\pm0.030$ & $ 0.010\pm0.030$ & $ 0.005\pm0.041$ & $ 0.011\pm0.030$ & $-0.039\pm0.030$ & $-0.018\pm0.030$ \\
16 &  S &  664 &  10.36 & $ 0.148\pm0.037$ &          $\dots$ & $-0.053\pm0.061$ &          $\dots$ &          $\dots$ &          $\dots$ \\
19 &  K & 1006 &   4.34 & $ 0.070\pm0.074$ &          $\dots$ &          $\dots$ &          $\dots$ &          $\dots$ &          $\dots$ \\
20 & Ca & 1517 &   6.11 & $-0.016\pm0.030$ & $-0.014\pm0.070$ & $ 0.006\pm0.030$ & $ 0.033\pm0.030$ & $-0.001\pm0.030$ & $ 0.016\pm0.030$ \\
21 & Sc & 1659 &   6.54 & $ 0.047\pm0.066$ & $ 0.073\pm0.048$ & $ 0.042\pm0.080$ & $ 0.028\pm0.030$ & $-0.061\pm0.041$ &          $\dots$ \\
22 & Ti & 1582 &   6.82 & $-0.025\pm0.030$ & $ 0.097\pm0.047$ & $ 0.017\pm0.030$ & $ 0.039\pm0.030$ & $-0.007\pm0.030$ & $ 0.004\pm0.030$ \\
23 &  V & 1429 &   6.74 & $-0.067\pm0.030$ &          $\dots$ & $-0.014\pm0.054$ & $ 0.014\pm0.030$ & $-0.020\pm0.030$ &          $\dots$ \\
24 & Cr & 1296 &   6.77 & $-0.020\pm0.030$ &          $\dots$ & $ 0.013\pm0.030$ & $ 0.006\pm0.030$ & $-0.005\pm0.030$ & $-0.013\pm0.030$ \\
25 & Mn & 1158 &   7.43 & $-0.085\pm0.039$ &          $\dots$ & $-0.015\pm0.056$ & $-0.023\pm0.031$ &          $\dots$ &          $\dots$ \\
27 & Co & 1352 &   7.86 & $-0.047\pm0.051$ & $-0.012\pm0.033$ & $-0.010\pm0.050$ & $ 0.001\pm0.030$ & $-0.041\pm0.042$ &          $\dots$ \\
28 & Ni & 1353 &   7.63 & $-0.023\pm0.030$ & $-0.007\pm0.030$ & $-0.008\pm0.031$ & $-0.010\pm0.030$ & $-0.035\pm0.030$ & $-0.020\pm0.031$ \\
29 & Cu & 1037 &   7.73 & $-0.070\pm0.067$ & $-0.032\pm0.062$ & $-0.058\pm0.063$ &          $\dots$ &          $\dots$ &          $\dots$ \\
30 & Zn &  726 &   9.39 & $ 0.055\pm0.042$ & $ 0.040\pm0.089$ & $-0.005\pm0.055$ &          $\dots$ &          $\dots$ & $-0.018\pm0.052$ \\
38 & Sr & 1464 &   5.70 & $-0.008\pm0.080$ &          $\dots$ &          $\dots$ &          $\dots$ &          $\dots$ &          $\dots$ \\
39 &  Y & 1659 &   6.38 & $ 0.077\pm0.060$ & $ 0.113\pm0.090$ &          $\dots$ &          $\dots$ &          $\dots$ & $ 0.010\pm0.047$ \\
40 & Zr & 1741 &   6.84 & $ 0.047\pm0.118$ &          $\dots$ &          $\dots$ &          $\dots$ &          $\dots$ &          $\dots$ \\
56 & Ba & 1455 &   5.21 & $ 0.050\pm0.032$ & $ 0.095\pm0.150$ &          $\dots$ &          $\dots$ &          $\dots$ & $ 0.066\pm0.080$ \\
58 & Ce & 1478 &   5.47 & $ 0.055\pm0.071$ & $ 0.156\pm0.100$ &          $\dots$ &          $\dots$ &          $\dots$ &          $\dots$ \\
60 & Nd & 1602 &   5.49 & $ 0.085\pm0.057$ & $ 0.133\pm0.090$ &          $\dots$ &          $\dots$ &          $\dots$ &          $\dots$ \\
63 & Eu & 1356 &   5.67 & $ 0.040\pm0.051$ & $ 0.086\pm0.056$ &          $\dots$ &          $\dots$ &          $\dots$ &          $\dots$ \\

\hline
\end{tabular}
\label{t:abund_tcond}
\end{table*}

\begin{table}
\caption{Average (star-to-star) elemental abundance ratios [X/Fe] (and 1-$\sigma$ scatter) of solar twins in the \citet[\citetalias{ramirez09}]{ramirez09} and \citet[\citetalias{melendez09:twins}]{melendez09:twins} studies.}
\centering
\begin{tabular}{rlrr}\hline\hline
Z & El. & \multicolumn{1}{c}{\citetalias{melendez09:twins}} & \multicolumn{1}{c}{\citetalias{ramirez09}} \\ \hline 
 6 &  C & $-0.048\pm0.042$ & $-0.065\pm0.033$ \\
 7 &  N & $-0.060\pm0.051$ &          $\dots$ \\
 8 &  O & $-0.033\pm0.038$ & $-0.015\pm0.028$ \\
11 & Na & $-0.024\pm0.024$ & $-0.052\pm0.027$ \\
12 & Mg & $-0.018\pm0.030$ &          $\dots$ \\
13 & Al & $ 0.033\pm0.025$ & $-0.012\pm0.053$ \\
14 & Si & $-0.030\pm0.020$ & $-0.012\pm0.019$ \\
15 &  P & $-0.051\pm0.063$ &          $\dots$ \\
16 &  S & $-0.056\pm0.042$ & $-0.036\pm0.028$ \\
19 &  K & $-0.013\pm0.020$ &          $\dots$ \\
20 & Ca & $-0.004\pm0.010$ & $ 0.024\pm0.019$ \\
21 & Sc & $ 0.018\pm0.031$ & $ 0.024\pm0.018$ \\
22 & Ti & $ 0.011\pm0.016$ & $ 0.028\pm0.031$ \\
23 &  V & $-0.009\pm0.019$ & $ 0.014\pm0.015$ \\
24 & Cr & $-0.011\pm0.012$ & $ 0.016\pm0.017$ \\
25 & Mn & $-0.029\pm0.018$ & $-0.018\pm0.021$ \\
27 & Co & $-0.011\pm0.028$ &          $\dots$ \\
28 & Ni & $-0.007\pm0.013$ & $-0.021\pm0.013$ \\
29 & Cu & $-0.021\pm0.023$ & $-0.025\pm0.033$ \\
30 & Zn & $-0.023\pm0.032$ & $-0.027\pm0.061$ \\
39 &  Y & $ 0.015\pm0.033$ & $ 0.024\pm0.043$ \\
40 & Zr & $ 0.023\pm0.027$ & $ 0.037\pm0.060$ \\
56 & Ba & $ 0.012\pm0.025$ & $ 0.070\pm0.055$ \\

\hline
\end{tabular}
\label{t:abund_tcond_rm09}
\end{table}

\begin{table}
\caption{Average (star-to-star) elemental abundance ratios (and 1-$\sigma$ scatter) of solar analogs from the six independent studies examined in this work. The second column gives the average abundance ratios for the purely spectroscopic papers only.}
\centering
\begin{tabular}{rlrr}\hline\hline
Z & El. & \multicolumn{1}{c}{average (all)} & \multicolumn{1}{c}{average (spec.)} \\ \hline 
 6 &  C & $-0.044\pm0.111$ &          $\dots$ \\
 8 &  O & $-0.033\pm0.071$ & $-0.008\pm0.038$ \\
11 & Na & $-0.048\pm0.020$ & $-0.048\pm0.020$ \\
12 & Mg & $ 0.002\pm0.026$ & $ 0.001\pm0.029$ \\
13 & Al & $ 0.005\pm0.027$ & $-0.003\pm0.024$ \\
14 & Si & $-0.000\pm0.023$ & $-0.012\pm0.020$ \\
16 &  S & $-0.053\pm0.061$ & $-0.053\pm0.061$ \\
20 & Ca & $ 0.007\pm0.020$ & $ 0.014\pm0.020$ \\
21 & Sc & $ 0.042\pm0.020$ & $ 0.030\pm0.020$ \\
22 & Ti & $ 0.012\pm0.032$ & $ 0.013\pm0.020$ \\
23 &  V & $-0.023\pm0.032$ & $-0.004\pm0.020$ \\
24 & Cr & $-0.004\pm0.020$ & $ 0.000\pm0.020$ \\
25 & Mn & $-0.042\pm0.030$ & $-0.021\pm0.020$ \\
27 & Co & $-0.016\pm0.020$ & $-0.013\pm0.020$ \\
28 & Ni & $-0.017\pm0.020$ & $-0.018\pm0.020$ \\
29 & Cu & $-0.045\pm0.020$ & $-0.058\pm0.063$ \\
30 & Zn & $-0.004\pm0.020$ & $-0.012\pm0.020$ \\
38 & Sr & $-0.008\pm0.080$ &          $\dots$ \\
39 &  Y & $ 0.047\pm0.041$ & $ 0.010\pm0.047$ \\
40 & Zr & $ 0.047\pm0.118$ &          $\dots$ \\
56 & Ba & $ 0.054\pm0.020$ & $ 0.066\pm0.080$ \\
58 & Ce & $ 0.089\pm0.067$ &          $\dots$ \\
60 & Nd & $ 0.099\pm0.031$ &          $\dots$ \\
63 & Eu & $ 0.061\pm0.032$ &          $\dots$ \\

\hline
\end{tabular}
\label{t:abund_tcond_avg}
\end{table}

Fig.~\ref{f:abund_tcond}c shows that the abundance data for solar analogs by \citetalias{takeda07:abundances} are consistent with the [X/Fe] vs.\ $\tc$ trend found in our previous papers. The most volatile elements (C, N, O) are not covered by \citetalias{takeda07:abundances} but the break in the $\tc$ trend at about $\tc\simeq1000$\,K, which is crucial for the interpretation of this abundance trend (Sect.~\ref{s:signature}), appears to be detected.

The abundance vs.\ $\tc$ trend obtained for the \citetalias{neves09} sample of solar twins is in excellent agreement with \citetalias{ramirez09} and \citetalias{melendez09:twins} (Fig.~\ref{f:abund_tcond}d). Note, however, that this comparison is limited to refractory elements since no volatiles were analyzed by \citetalias{neves09}. In any case, both the overall effect detected by \citetalias{melendez09:twins} and  \citetalias{ramirez09} and the value of the [X/Fe] vs.\ $\tc$ slope for $\tc>900$\,K are in good agreement with the \citetalias{neves09} data for solar analogs. Moreover, $\simeq2-4$ of the 19 solar analogs from \citetalias{neves09} show a [X/Fe] vs.~$\tc$ slope close to zero (the uncertainties in the abundance ratios of individual stars prevent us from determining this number more accurately). This frequency of stars which do not follow the $\tc$ trend (10--20\,\%) is also in good agreement with the $\simeq15$\,\% frequency found by both \citetalias{ramirez09} and \citetalias{melendez09:twins}. A previous investigation of $\tc$ trends by the same group has been published before \citep{ecuvillon06}. This previous work was unable to reach strong conclusions regarding $\tc$ trends, probably because no investigation of solar twin stars alone was made. This study also adopted solar abundances from other works, mainly \cite{anders89}, instead of performing a strictly differential analysis using their own solar spectra. We have shown that both the sample restriction to solar twins and strictly differential analysis are necessary to detect the small $\tc$ trend in the abundances. The fact that this group \citep{ecuvillon06,neves09} has now been able to detect the abundance signature using our approach \citep{gonzalez-hernandez10} confirms this statement.

In the $\tc$ range between about 900 and 1600\,K, the agreement between the \citetalias{ramirez09} [X/Fe] vs.\ $\tc$ trend and that given by the \citetalias{gonzalez10} solar twin data is good, albeit a very small offset in [Fe/H] of at most 0.02\,dex (Fig.~\ref{f:abund_tcond}e). The two elements with the highest $\tc$ in the \citetalias{gonzalez10} work are Al and Sc. The average Al abundance from \citetalias{gonzalez10} is in good agreement with that by \citetalias{ramirez09} and they are both within about 1-$\sigma$ of the mean [X/Fe] vs.\ $\tc$ trend. This is not the case for the Sc abundance by \citetalias{gonzalez10}. As explained in Sect.~\ref{s:samples}, this discrepancy may be due to the use of \ion{Sc}{ii} lines in the \citetalias{gonzalez10} work as opposed to \ion{Sc}{i} lines as in \citetalias{ramirez09}. The latter were shown to reveal less star-to-star scatter for the solar twins and therefore they are expected to be more reliable in strictly differential analysis. With the exception of the Sc abundances, the $\tc$ trend in the \citetalias{gonzalez10} data is consistent with that in \citetalias{ramirez09} and \citetalias{melendez09:twins}.

Although the average [X/Fe] vs.\ $\tc$ trend obtained for the \citetalias{bensby11} data is in agreement with the \citetalias{ramirez09} result within 1-$\sigma$, it is clear that the $\tc$ trend of refractories is slightly more shallow in the \citetalias{bensby11} case (Fig.~\ref{f:abund_tcond}f). The difference in the mean abundance of refractory and volatile elements, weighted by the errors, is only about 0.01\,dex, but the sign of such difference is in agreement with all other studies presented thus far in this paper as well as \citetalias{melendez09:twins} and \citetalias{ramirez09}.

\begin{figure}
\includegraphics[width=9.2cm,bb=75 375 560 700]
{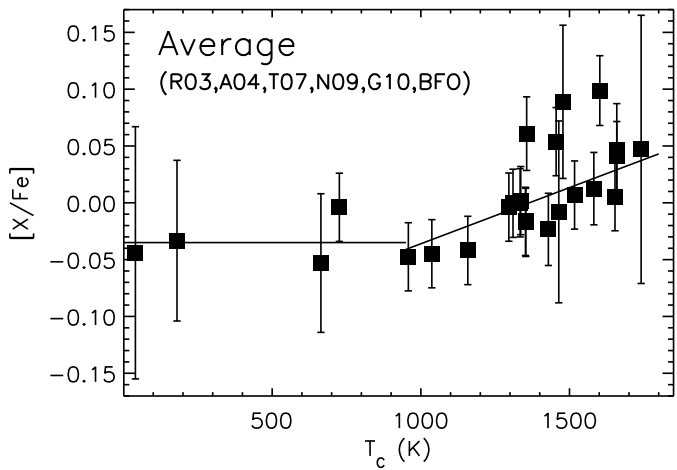}
\caption{Abundance ratios relative to iron as a function of condensation temperature for all samples of solar twins and analogs investigated in this paper (weighted average and 1-$\sigma$ scatter of mean abundance ratios from different studies). Solid lines represent the average [X/Fe] versus $\tc$ relation for solar twin stars obtained by \cite{ramirez09}.}
\label{f:all_tcond}
\end{figure}

In Fig.~\ref{f:all_tcond} we show an average [X/Fe] vs.\ $\tc$ relation for all six samples \citepalias{reddy03,allende04:s4n,takeda07:abundances,neves09,gonzalez10,bensby11} combined. For the elements present in more than one study we computed the weighted mean of the individually averaged results. We excluded the \citetalias{reddy03} results for the volatiles because they are likely affected by systematic errors in the temperature scale, measurement of equivalent widths, and/or chemical evolution effects, as explained above. We also excluded the K abundances from this study because of their large internal error ($\simeq0.07$\,dex) and the fact that they are based on the analysis of only one very strong feature ($EW\simeq170$\,m\AA), which could be affected by continuum placement and saturation. The impact of the mean Sc abundance from the \citetalias{gonzalez10} solar analogs, which we suspect to be slightly off, on this result is minor but we excluded it for consistency. The agreement of this average [X/Fe] vs.~$\tc$ trend with the \citetalias{melendez09:twins} and \citetalias{ramirez09} results is good, but we emphasize the fact that this is an heterogeneous compilation of abundances and that the different samples used for this compilation exhibit a slightly different behavior when analyzed separately. In fact, we argue that because the data used to make this figure are heterogeneous and still show a clear abundance vs.\ $\tc$ correlation, it strengthens the findings of \citetalias{melendez09:twins} and \citetalias{ramirez09}. The [X/Fe] abundance ratios for the average of the six samples combined are given in Table~\ref{t:abund_tcond_avg}. 

\begin{figure}
\includegraphics[width=9.2cm,bb=75 375 560 700]
{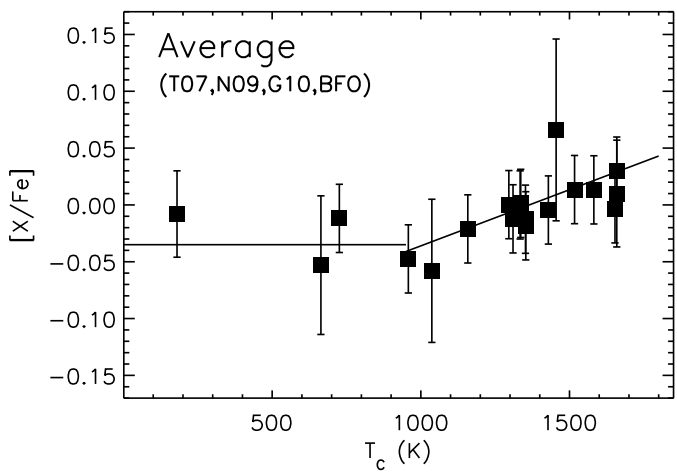}
\caption{As in Fig.~\ref{f:all_tcond} for the four more recent papers, which use only the spectra (and not photometry and/or trigonometric parallaxes) to determine stellar parameters.}
\label{f:all_tcond_recent}
\end{figure}

Inspection of Fig.~\ref{f:abund_tcond} clearly reveals that the results obtained for the \citetalias{reddy03} and \citetalias{allende04:s4n} samples are much more noisy than those given in the more recent papers. We note that these two older studies are not purely spectroscopic. They both use photometric data to estimate effective temperatures and trigonometric parallaxes to determine the surface gravities, therefore they are not strictly differential. This is not the case for the more recent papers, which use only the spectra to determine the stellar parameters. To make sure that Fig.~\ref{f:all_tcond} is not biased by these two older studies, we re-computed the average abundances but only for \citetalias{takeda07:abundances}, \citetalias{neves09}, \citetalias{gonzalez10}, and \citetalias{bensby11}. The result is shown in Fig.~\ref{f:all_tcond_recent}, where it can be seen that the conclusion is not altered by excluding the more noise data. The last column of Table~\ref{t:abund_tcond_avg} gives the average abundance ratios plotted in Fig.~\ref{f:all_tcond_recent}.

None of the abundance studies scrutinized in this paper disagree with the \citetalias{melendez09:twins} and \citetalias{ramirez09} results. In particular, with the exception of the \citetalias{reddy03} results, which we suspect to be in error for the volatiles, none of them suggest a significant enhancement of volatiles with respect to refractories. Although very good agreement with our previous results is only found for two of the six studies, the other four are still consistent with them.

The internal errors in the abundance ratios of the six studies used in this paper are expected to be in the 0.04 to 0.10 dex range, which is similar to the amplitude of the effect detected by \citetalias{melendez09:twins} and \citetalias{ramirez09}. The latter, on the other hand, reported abundance ratios with a precision of about 0.01 and 0.03\,dex, respectively, which were achieved thanks to the extremely high quality and homogeneity of the data, as well as the high internal consistency of the analysis. Therefore, it is not surprising that the $\tc$ trends observed by \citetalias{melendez09:twins} and \citetalias{ramirez09} were more difficult to detect in the six studies compiled for this paper.

\section{A signature of terrestrial planet formation?} \label{s:signature}

As explained by \citetalias{melendez09:twins} and also \citetalias{ramirez09}, the solar abundance anomaly (i.e., the deficiency of refractory elements relative to volatiles in the Sun as compared to solar twins) can be interpreted as a signature of the formation of planets, and perhaps more specifically terrestrial planets, in the solar system. They speculate that the accretion of chemically fractionated material into the solar convective envelope during the formation of the solar system made its chemical composition pattern depleted in refractories as these elements were retained by the planets. The fact that the total mass of refractories in the terrestrial planets of the solar system today is of the same order of magnitude as the mean observed difference between refractories and volatiles in the Sun relative to solar twins supports the hypothesis of a \textit{terrestrial} planet signature, as well as the observed abundance pattern of meteorites, which mirrors the solar abundance anomaly \citep{alexander01,ciesla08}.

Fig.~\ref{f:all_tcond} shows that the abundance ratios are a roughly linear increasing function of condensation temperature for $\tc\gtrsim900$\,K but they become nearly constant for $\tc\lesssim900$\,K. This suggests that the process responsible for the observed trend affects only elements with $\tc\gtrsim900$\,K.\footnote{Note that our abundance ratios are all relative to Fe, which, within our ``planet formation signature'' scenario, is depleted in the solar photosphere. Had we chosen a volatile element as reference, abundance ratios of $\tc\lesssim900$\,K elements would be nearly constant at about zero.} These high temperatures are only found in the inner parts ($\lesssim$2\,AU) of protoplanetary disks, regions which, at least in the solar system, are inhabited by terrestrial planets. Moreover, \citetalias{melendez09:twins} analyzed the abundance vs.\ $\tc$ trends of about 10 solar analogs with and without detected giant gas planets. They found that those stars hosting hot Jupiters all behave like the majority of solar twin stars, i.e., they have a chemical composition slightly different from solar. More striking is the fact that the fraction of stars which closely resemble the Sun in their chemical composition is about 70\,\% if the star does not have a close-in giant planet. Within our interpretation of the $\tc$ trend as a planet signature, we are led to conclude that the presence of hot Jupiters prevents the formation of terrestrial planets. One could also speculate that in those systems, the smaller planets have already been accreted by the host star, thus removing the initially imprinted abundance signature.

In order for this proposed terrestrial planet signature to be imprinted in the chemical composition of the Sun, however, a quite short time-scale for the formation of the solar convective zone and/or an unusually long lived protoplanetary disk are required (\citetalias{melendez09:twins}, \citealt{nordlund09}, \citealt{gustafsson10}). Standard hydrostatic models for the early evolution of the solar interior suggest that it took about 30\,Myr for the convective zone to reach its present size \citep[e.g.,][]{dantona94} while age determinations of observed disks around young stars suggest that they dissipate in about 10\,Myr, at most \citep[e.g.,][]{mamajek09,meyer09,fedele10}. If this were the case, the chemically fractionated material would have been mixed efficiently in an early massive convective zone and the signature of planet formation would have been diluted. Hydrodynamic models of early stellar evolution, however, suggest that the Sun was never fully convective and that the internal structure of the early Sun ($\simeq1$\,Myr) was not very different from that of the current Sun (\citealt{wuchterl03}). Within this latter scenario, the abundance signature can be effectively imprinted, even if the protoplanetary disk was short-lived. Another alternative for solving this problem involves the inclusion of episodic accretion in early stellar evolution calculations, which tends to decrease the size of the convective envelope \citep{baraffe09,baraffe10}.

\begin{figure}
\includegraphics[width=9.5cm,bb=65 370 340 548]
{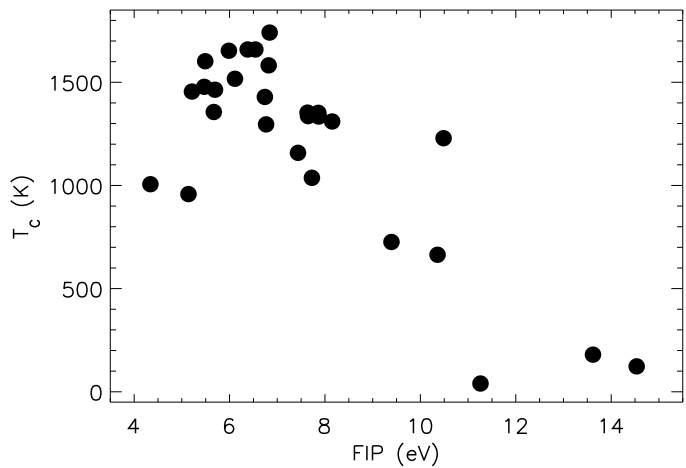}
\caption{Condensation temperature ($\tc$) versus first ionization potential (FIP) for the elements listed in Tables~\ref{t:abund_tcond} and \ref{t:abund_tcond_rm09}.}
\label{f:tc_fip}
\end{figure}

\begin{figure}
\includegraphics[width=9.7cm,bb=83 380 340 815]
{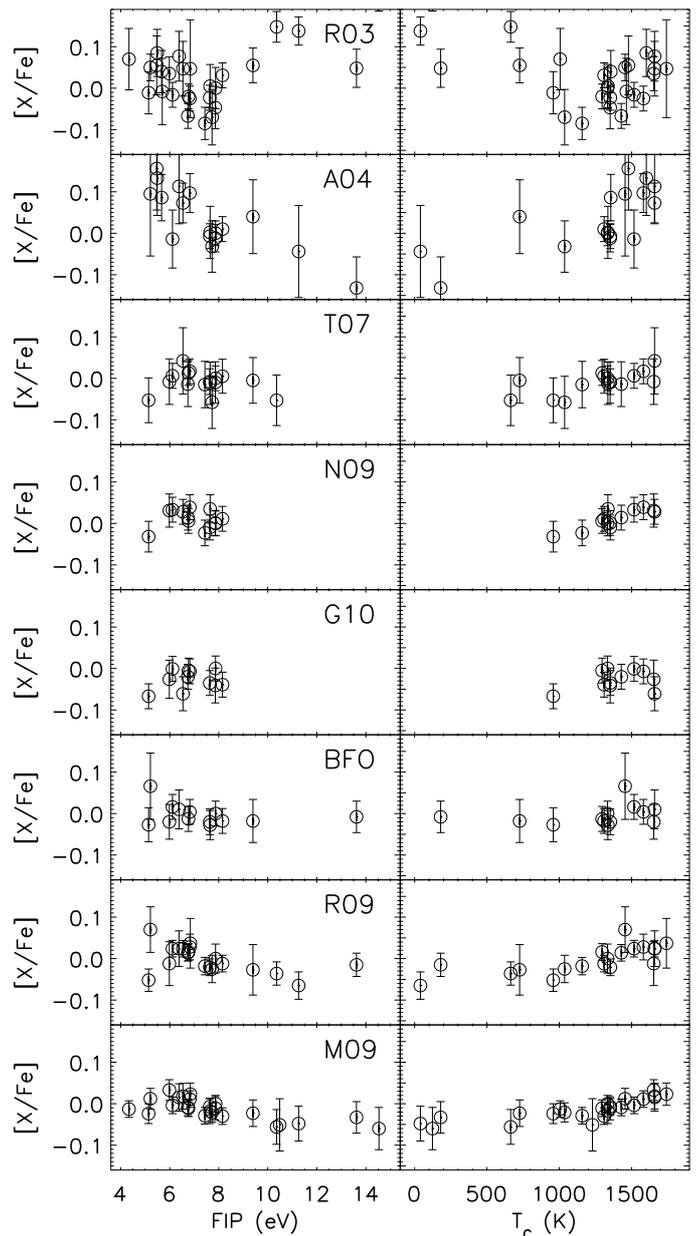}
\caption{Comparison of the mean abundance ratio [X/Fe] versus first ionization potential (FIP) relation (left panels) with the [X/Fe] versus condensation temperature ($\tc$) relation (right panels) for each of the samples studied in this paper (top six panels), and the \citet[\citetalias{ramirez09}]{ramirez09} and \citet[\citetalias{melendez09:twins}]{melendez09:twins} studies (bottom two panels).}
\label{f:fip}
\end{figure}

The condensation temperatures of elements correlate very well with their first ionization potentials (FIP), as shown in Fig.~\ref{f:tc_fip} for the elements analyzed in this paper. One may therefore question whether the abundance trend with $\tc$ that we attribute to planet formation is instead due to an FIP-related effect. The abundances of low FIP elements in the (average quiet) upper solar atmosphere are found to be about 4 times greater than those observed in the photosphere \citep[e.g.,][]{feldman92}. This so-called ``FIP effect'' is attributed to the acceleration of low FIP ions from the chromosphere into higher layers, a process probably driven by magnetic fields \citep[e.g.,][]{henoux98}. The FIP effect is not expected to affect the photospheric abundances. Nevertheless, we compared the significance of the correlation between [X/Fe] and FIP with that corresponding to $\tc$ for the samples analyzed in this paper as well as those from the \citetalias{melendez09:twins} and \citetalias{ramirez09} studies. The results are given in Fig.~\ref{f:fip} and Tables~\ref{t:spearman} and \ref{t:spearman_refonly}.

\begin{table}
\caption{Spearman correlation coefficient $r_S$ and percent probability of correlation occurring by chance ($P_S$) for the samples shown in Fig.~\ref{f:fip}}
\centering
\begin{tabular}{l|rr|rr}\hline\hline
Sample & $r_S$ & $P_S$ & $r_S$ & $P_S$  \\
       & \multicolumn{2}{c|}{(vs.\ FIP)} & \multicolumn{2}{c}{(vs.\ $\tc$)} \\ \hline
R03    & $ 0.125$ & 54.8 & $-0.138$ & 51.5 \\
A04    & $-0.654$ &  1.5 & $ 0.648$ &  1.4 \\
T07    & $-0.090$ & 55.9 & $ 0.399$ & 21.1 \\
N09    & $-0.093$ & 61.9 & $ 0.576$ &  7.9 \\
G10    & $ 0.111$ & 60.7 & $ 0.063$ & 63.9 \\
BFO    & $-0.164$ & 50.6 & $ 0.278$ & 38.2 \\
R09    & $-0.433$ &  9.7 & $ 0.672$ &  0.8 \\
M09    & $-0.571$ &  1.3 & $ 0.736$ &  0.1 \\ \hline
\end{tabular}
\label{t:spearman}
\end{table}

\begin{table}
\caption{Spearman correlation coefficient $r_S$, percent probability of correlation occurring by chance ($P_S$), and significance of the [X/Fe] vs.\ FIP or $\tc$ slope for the $\tc>900$\,K data points of the samples shown in Fig.~\ref{f:fip}.}
\centering
\begin{tabular}{l|rrc|rrc}\hline\hline
Sample & $r_S$ & $P_S$ &  slope$/\sigma$ & $r_S$ & $P_S$ & slope$/\sigma$ \\
       & \multicolumn{3}{c|}{(vs.\ FIP)} & \multicolumn{3}{c}{(vs.\ $\tc$)} \\ \hline
R03    & $-0.397$ & 10.0 & $-2.1$ &  0.380 & 13.5 & 1.6 \\
A04    & $-0.616$ &  3.8 & $-2.9$ &  0.626 &  2.8 & 2.6 \\
T07    & $+0.024$ & 58.7 & $+0.1$ &  0.386 & 25.1 & 1.2 \\
N09    & $-0.093$ & 61.9 & $-0.3$ &  0.576 &  7.9 & 2.0 \\
G10    & $+0.111$ & 60.7 & $+0.6$ &  0.063 & 63.9 & 1.0 \\
BFO    & $-0.154$ & 55.5 & $-0.5$ &  0.322 & 36.3 & 0.9 \\
R09    & $-0.222$ & 46.2 & $-1.0$ &  0.588 &  5.1 & 2.9 \\
M09    & $-0.370$ & 17.4 & $-1.0$ &  0.682 &  0.6 & 2.6 \\ \hline
\end{tabular}
\label{t:spearman_refonly}
\end{table}

In most cases, the Spearman correlation coefficient (in absolute value) is larger for the $\tc$ trend than the FIP trend. For the \citetalias{reddy03} and \citetalias{allende04:s4n} samples, these values are similar although both correlations (FIP and $\tc$) are much more significant in the \citetalias{allende04:s4n} case. The result for the \citetalias{gonzalez10} data suggests a lack of correlation with both FIP and $\tc$. However, this is mainly due to their Sc abundances (see below for a quantitative statement), which we suspect to have a small but significant error in its zero point (Sect.~\ref{s:samples}). In four of the samples listed in Table~\ref{t:spearman} the $\tc$ trend has a correlation coefficient greater than 0.5. Only two samples show similarly high correlation coefficients for the FIP trends. In none of the cases investigated here is the correlation with FIP much more significant than that with $\tc$. This situation is made even more clear when only the abundances of refractory elements are analyzed (Table~\ref{t:spearman_refonly}). The correlation coefficients for the $\tc$ trends suggest a similar or higher significance compared to the FIP trends for all samples except \citetalias{gonzalez10}, unless we exclude their Sc abundances that have a suspected small offset, in which case $r_S=0.054$ for FIP and $r_S=0.278$ for $\tc$. For four cases the FIP trend is not significant (\citetalias{takeda07:abundances}, \citetalias{neves09}, \citetalias{gonzalez10}, and \citetalias{bensby11}).  We also note that a linear fit to the abundance vs. $\tc$ trend of refractories is, for the majority of cases, more significant than the FIP trend. For example, the slope with $\tc$ for the \citetalias{neves09} sample is significant at the 2.0\,$\sigma$ level whereas that corresponding to the FIP trend is not significant at all. For the \citetalias{gonzalez10} results excluding Sc we find that the $\tc$ slope divided by its error is about 5 times larger (in absolute value) than that corresponding to the FIP trend. Thus, we find that, statistically, the [X/Fe] abundance ratios correlate better with $\tc$ than FIP. Note that this is particularly true for the \citetalias{melendez09:twins} and \citetalias{ramirez09} results, which are the most reliable due to the particularly high abundance accuracy.

Finally, we should also note that, for all studies in which this element is included (the only exception being \citetalias{allende04:s4n}), Na, which has one of the lowest FIP (5.139~eV) but not a particularly high $\tc$ (958\,K), fits very well the abundance vs.~$\tc$ trend but reveals itself as an outlier of the abundance vs.~FIP correlation. A similar behavior is exhibited by K but, as noted in Sect~\ref{s:tc_trend}, its abundance is not as reliable.

\section{Future directions}

The literature data for solar analogs analyzed in this paper support the observational findings of \citetalias{melendez09:twins} and \citetalias{ramirez09}. The uncertainties in the abundances of individual stars are significantly larger than those by \citetalias{melendez09:twins} and \citetalias{ramirez09} and it is therefore not possible to establish with confidence which fraction of stars have abundances extremely close to solar (except for \citetalias{neves09} for which we find a value of 10 to 20\,\%). According to our interpretation, stars with no deficiency of refractories relative to volatiles (or overabundance of volatiles) relative to the Sun should also host terrestrial planets. \citetalias{melendez09:twins} and \citetalias{ramirez09} were able to determine a frequency of about 15\,\% for these objects.

\subsection{Metallicity effects}

From studies of extrasolar giant gas planets, it has been well established that the frequency of planet-hosts is roughly proportional to $10^{2\feh}$ \citep[e.g.,][]{marcy05}. If we assume that the relative frequency of planet hosts is independent of whether the planets are gas giants or terrestrial planets, the fraction of terrestrial planets at $\feh\simeq-0.2$ should be about 6\,\% while that at $\feh\simeq+0.2$ it should be about 40\,\% based on the above-mentioned 15\,\% frequency at $\feh=0$. For a sample of about 20 stars, this means that at $\feh\simeq-0.2$ the number of stars which would closely resemble the Sun in their chemical composition is 1. At $\feh\simeq0.0$ the corresponding number is 3, as in \citetalias{ramirez09}. In both cases such a low number makes these objects an obvious minority and therefore the solar abundance anomaly manifests itself clearly. Following this line of reasoning, at $\feh\simeq+0.2$ the number of stars without $\tc$ trends in the abundances should be about 8, a number that is comparable to that of those stars which show the $\tc$ trend (12). If the hypothesis of planet formation signature is correct and the observed frequency of gas giant planet hosts as a function of $\feh$ applies also to the planets responsible for the abundance signature we have uncovered, the $\tc$ trend with (average) abundance ratios should progressively disappear at higher $\feh$. The signature in this case has been washed out by taking the average of a similar fraction of stars which show and do not show the $\tc$ trend. If sufficient precision is achieved, it may be possible to distinguish between metal-rich stars that show and do not show the planet signature from a bimodal distribution of $\tc$ slopes for refractory elements.

\begin{figure}
\includegraphics[width=9.2cm,bb=75 375 340 685]
{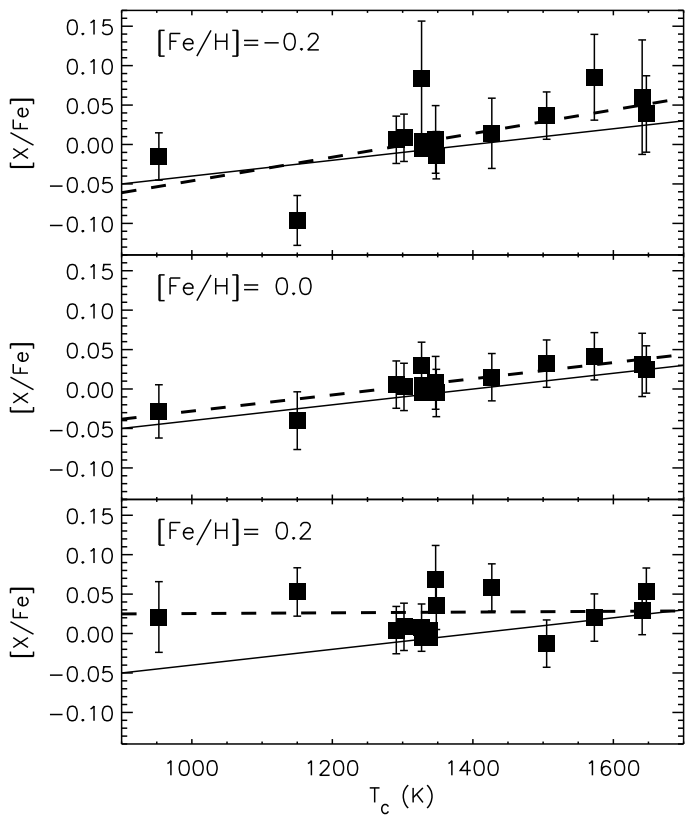}
\caption{Upper panel: Average [X/Fe] versus $\tc$ relation for refractory elements in solar analogs with $\feh=-0.2\pm0.1$. Data are from \cite{neves09}. The dashed line is a linear fit to the data while the solid line, which corresponds to the average relation for solar twins found by \cite{ramirez09}, has been drawn for comparison. Middle panel: as in the upper panel for $\feh=+0.0\pm0.1$. Lower panel: as in the upper panel for $\feh=+0.2\pm0.1$.}
\label{f:neves_tcond_feh}
\end{figure}

Interestingly, the behavior of the average abundance trend of solar analogs with $\tc$ as a function of $\feh$ described above is in good agreement with the \citetalias{neves09} data (Fig.~\ref{f:neves_tcond_feh}). We perform this test on this sample only because it shows the most clear and a very significant $\tc$ trend for refractory elements. Fig.~\ref{f:neves_tcond_feh} shows that the solar abundance anomaly is observed both at $\feh\simeq-0.2$ and $\feh\simeq0.0$. This is expected according to our interpretation because the fraction of stars with abundances extremely similar to solar (and therefore potential terrestrial planet hosts) is very small in both cases. At $\feh\simeq+0.2$ the fraction of terrestrial planet hosts is comparable to that of stars which do not host those planets and therefore the signature is no longer evident. We also expect the scatter to be larger in the metal-rich case because of the even mixture of stars that have different chemical compositions. Indeed, the weighted 1-$\sigma$ scatter of the average abundance ratios relative to the linear fits shown in Fig.~\ref{f:neves_tcond_feh} increases from about 0.012\,dex at $\feh=+0.0$ to 0.024\,dex at $\feh=+0.2$. We note, however, that the corresponding scatter at $\feh=-0.2$ is larger (0.033\,dex) than that of the two more metal-rich samples.

In addition to the result described above, \citetalias{ramirez09} found that the fraction of stars showing the planet signature in the abundances increases with $\feh$ while a hint of a bimodal distribution for the $\tc$ slopes of refractory elements was also observed (their Fig.~4). Although encouraging, these results rely on the precision of abundance ratios relative to solar for stars which are significantly more metal-rich (or more metal-poor in the \citetalias{neves09} case) than the Sun and therefore small systematic errors and/or chemical evolution may affect the abundance determinations. It is crucial that we repeat the exercise of looking for abundance trends with condensation temperature but using large samples ($\simeq50$) of metal-rich (or metal-poor) twin stars and analyzing them with respect to a well known metal-rich (or metal-poor) reference star.

\subsection{Effective temperature effects}

Our proposed planet signature in the abundances results from the accretion and subsequent mixing of chemically fractionated material into the stellar convective zone. Thus, the magnitude of the effect depends on the size of the convective envelope at the time of protoplanetary disk accretion. A massive envelope would dilute any chemical peculiarities of the accreted material. It is well established that the mass of the convective envelope reduces with effective temperature. In F-dwarfs at $\teff\simeq6250$\,K, the total mass of the convective envelope is $0.004\,M_\odot$, a figure that is significantly smaller than that corresponding to the Sun \citep[about $0.02\,M_\odot$;][]{pinsonneault01}. This implies that the chemical signature, if due to the accretion from a protoplanetary disk of similar mass as the one that gave origin to our solar system, should be enhanced by a factor of $\simeq5$ in mass, everything else being equal. This corresponds, on the logarithmic abundance ratio scale, to a maximum difference in refractory and volatile elements of 0.30\,dex instead of 0.08\,dex for the solar case. The samples studied in this paper do not exhibit such behavior. Note, however, that the analysis of these warmer stars is much more affected by systematic errors such as those due to departures from the local-thermodynamic equilibrium assumption as well as surface inhomogeneities \citep[e.g.,][]{asplund05:review}. Also, their relatively large projected rotational velocities and macroturbulent velocities make the identification of line blends more difficult and consequently the line profile fitting or equivalent width measurement less accurate. As with the metal-rich example described above, it is crucial that we analyze very high quality spectra of a large sample ($\simeq50$) of twin F-dwarfs with respect to a well-known object of that class in order to search for the abundance patterns found for solar twins stars, if indeed they are there.

\subsection{Recent developments}

During the review process of this paper, the two suggestions described above regarding the impact of metallicity and effective temperature on the abundance vs.\ $\tc$ trends have been tested by \citetalias{gonzalez10} but using a different approach, that of weighting the results according to the distance between stars in the stellar parameter space. The method that we use is more direct although it remains to be proven whether it is more reliable. The size of the samples that we propose to analyze (about 50 stars in each case) should allow us to make quantitative tests of the planet signature hypothesis.

\begin{figure}
\includegraphics[width=9.5cm,bb=80 375 560 700]
{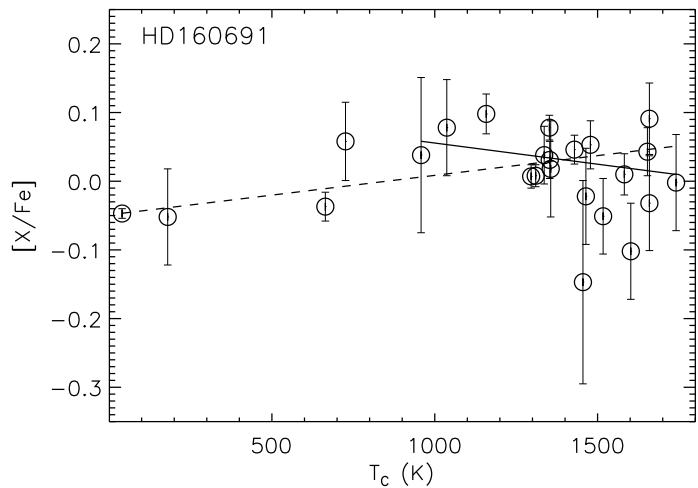}
\caption{Abundance ratios [X/Fe] vs.~condensation temperature $\tc$ for the super-Earth planet host HD\,160691. Abundance data are from \cite{gonzalez-hernandez10}. The dashed line is a linear fit to all elements; the solid line is a linear fit to refractories ($\tc>900$\,K) only, as was done by \citetalias{ramirez09} to investigate the planet signature interpretation. Volatiles must be excluded from this discussion because their abundances are not as reliable beyond the solar twin regime.}
\label{f:gh_abund_hd160691}
\end{figure}

Also during the review process of this paper, \cite{gonzalez-hernandez10} published a search for our proposed planet signatures in the abundances that they derived for a sample of solar analogs. Most of their sample stars and data employed are included in \citetalias{neves09}. They find good agreement between their abundance vs.~$\tc$ trends for solar twins and the results by \citetalias{melendez09:twins}. They claim, however, that their abundance trends for metal-rich solar analogs are incompatible with those by \citetalias{ramirez09}. In particular, they discuss their results for two stars known to host super-Earth like planets (HD\,160691 and HD\,1461). They find positive [X/Fe] vs.~$\tc$ slopes for these two stars, which in our interpretation would suggest the absence of rocky planets. In order to understand the apparent discrepancy with \citetalias{ramirez09}, in Fig.~\ref{f:gh_abund_hd160691} we show and examine the [X/Fe] vs.~$\tc$ trend obtained by \citeauthor{gonzalez-hernandez10} for one of these two super-Earth planet hosts (HD\,160691). The situation for HD\,1461 is very similar, as is that for the majority of the other super-solar metallicity stars discussed in that paper.

The $\tc$ slopes they use to examine our planet formation interpretation include both refractory and volatile elements. The two stars in question are both metal-rich. One must be careful when interpreting these abundance results beyond the realm of solar twins. Chemical evolution is likely affecting the abundances of the volatile elements C and O much more than the majority of refractories at these high metallicities. The other two volatile elements in their work are S and Zn, both of which can be very sensitive to errors in stellar parameters. In \citetalias{ramirez09} we excluded these elements from the discussion of $\tc$ slopes because the trend for refractories is more robust. Note also that, as we have pointed out before as being one of the most important characteristics of the [X/Fe] vs.~$\tc$ relation, the overall $\tc$ trend is not linear but rather shows a break at $\tc\simeq1000$\,K.

Fig.~\ref{f:gh_abund_hd160691} shows that the $\tc$ slope for HD\,160691 is positive if one includes all elements (dashed line).\footnote{The value that we derive for this slope is $+0.057\times10^{-3}$\,dex\,K$^{-1}$, which is slightly different from that given by \cite{gonzalez-hernandez10} because they mistakenly considered Sr a volatile element, placing it at $\tc\simeq450$\,K rather than at $\tc=1464$\,K, as provided by \cite{lodders03}. Note that this is also true for all the $\tc$ slopes presented in that paper.} As stated above, however, the abundances of volatiles are not as reliable as the trend for refractories, which is the reason why \citetalias{ramirez09} interpretation is based on $\tc$ slopes for $\tc>900$\,K elements only. Interestingly, a linear fit to the refractory elements only ($\tc>900$\,K) reveals a negative slope of $-0.062\times10^{-3}$\,dex\,K$^{-1}$ for HD\,160691 (solid line in Fig.~\ref{f:gh_abund_hd160691}) and also a negative slope of $-0.067\times10^{-3}$\,dex\,K$^{-1}$ for HD\,1461 instead of the positive slopes obtained by \citeauthor{gonzalez-hernandez10}

\begin{figure}
\includegraphics[width=9.5cm,bb=80 375 560 700]
{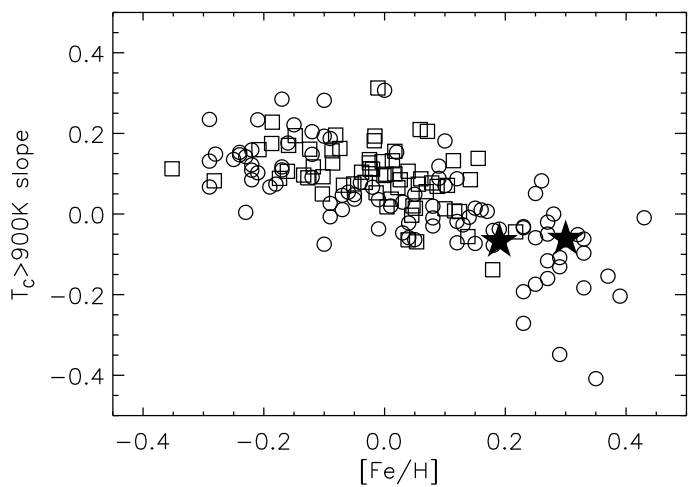}
\caption{Slopes of the [X/Fe] vs.~$\tc$ relation for refractory elements only ($\tc>900$\,K) as derived by \citetalias{ramirez09} (squares) and \cite{gonzalez-hernandez10} (circles). Two super-Earth planet hosts are shown with the star symbols.}
\label{f:gh_slope_params_consistent}
\end{figure}

In Fig.~\ref{f:gh_slope_params_consistent} we show the distribution of $\tc>900$\,K slopes for both the \citetalias{ramirez09} and \cite{gonzalez-hernandez10} samples (we derived the slopes for the latter using the same procedure used in \citetalias{ramirez09} and the abundance data provided in their paper). There is good agreement in the distribution of $\tc>900$\,K slopes of these two studies even though the \citeauthor{gonzalez-hernandez10} sample extends to higher metallicities. Moreover, the two super-Earth planet hosts, shown with the star symbols in Fig.~\ref{f:gh_slope_params_consistent} fall in the midst of the distribution of slopes for metal-rich stars.

The hint of a bimodal distribution of $\tc>900$\,K slopes at high metallicity speculated by \citetalias{ramirez09} is not confirmed by the \cite{gonzalez-hernandez10} data. Instead, beyond $\feh\simeq0.15$, almost all stars show near zero or negative slope. Within the planet formation signature in the abundances scenario, this suggests that all stars with metallicities higher than $\feh\simeq+0.15$ are terrestrial planet hosts. In particular, the two super-Earth planet hosts HD\,160691 and HD\,1461 both reveal the signature of terrestrial planets in their abundance pattern. We thus conclude that, when examined properly, the abundance vs. $\tc$ trends of these two super-Earth planet hosts are fully consistent with \citetalias{ramirez09} results and interpretation.

\section{Summary}

We have compiled abundance data for six independent spectroscopic surveys (\citealt{reddy03,allende04:s4n,takeda07:abundances,neves09,gonzalez10}; and \citeauthor{bensby11}, in preparation) and computed mean abundance ratios [X/Fe] for solar twins and analogs in each of them. With the exception of one study, we find that [X/Fe] correlates with the condensation temperature of the elements in a manner similar to that found by \citetalias{melendez09:twins} and \citetalias{ramirez09}. We show that it is very likely that the one study \citep{reddy03} that disagrees with the \citetalias{melendez09:twins} and \citetalias{ramirez09} results suffers from systematic errors that affect mostly the abundances of volatile elements, which are the only discrepant ones. Quantitatively, the agreement between the $\tc$ trends derived by us in our previous work and those from the literature is good for two of the six studies analyzed, while it is still consistent for the other four.

When the abundances from the six studies used in this paper are combined without attempting to homogenize the abundance results in any way, the agreement with \citetalias{melendez09:twins} and \citetalias{ramirez09} is also good. We argue that the heterogeneous nature of this compilation makes the finding of a good agreement even more robust. All six studies are completely independent and employ largely different spectra, model atmospheres, spectrum synthesis codes, linelists, etc.

We have investigated whether the abundance trends that we find with condensation temperature are due to an underlying correlation with the first ionization potential of the elements. Statistical tests show that condensation temperature is the driving parameter in these correlations.

The abundance trends with condensation temperature found for solar twins by \citetalias{melendez09:twins} and \citetalias{ramirez09} and shown to be consistent with our survey of literature data are speculated to be due to the formation of terrestrial planets in the Sun and a minority of solar twins. We provide suggestions to test specific predictions of this hypothesis.

\begin{acknowledgements}
J.\,M.\ would like to acknowledge support from Funda\c{c}\~ao para a Ci\^encia e a Tecnologia (FCT, Portugal) in the form of a grant (PTDC/CTE-AST/098528/2008) and a Ci\^encia 2007 fellowship. We thank G.~Gonzalez for sending us the abundance data from the \citetalias{gonzalez10} paper before publication.
\end{acknowledgements}

\bibliographystyle{aa}

\begin{thebibliography}{61}
\expandafter\ifx\csname natexlab\endcsname\relax\def\natexlab#1{#1}\fi

\bibitem[{{Alexander} {et~al.}(2001){Alexander}, {Boss}, \&
  {Carlson}}]{alexander01}
{Alexander}, C.~M.~O., {Boss}, A.~P., \& {Carlson}, R.~W. 2001, Science, 293,
  64

\bibitem[{{Allende~Prieto} {et~al.}(2004){Allende~Prieto}, {Barklem},
  {Lambert}, \& {Cunha}}]{allende04:s4n}
{Allende~Prieto}, C., {Barklem}, P.~S., {Lambert}, D.~L., \& {Cunha}, K. 2004,
  \aap, 420, 183

\bibitem[{{Alonso} {et~al.}(1996){Alonso}, {Arribas}, \&
  {Martinez-Roger}}]{alonso96}
{Alonso}, A., {Arribas}, S., \& {Martinez-Roger}, C. 1996, \aap, 313, 873

\bibitem[{{Anders} \& {Grevesse}(1989)}]{anders89}
{Anders}, E. \& {Grevesse}, N. 1989, \gca, 53, 197

\bibitem[{{Asplund}(2005)}]{asplund05:review}
{Asplund}, M. 2005, \araa, 43, 481

\bibitem[{{Baraffe} \& {Chabrier}(2010)}]{baraffe10}
{Baraffe}, I. \& {Chabrier}, G. 2010, \aap, submitted

\bibitem[{{Baraffe} {et~al.}(2009){Baraffe}, {Chabrier}, \&
  {Gallardo}}]{baraffe09}
{Baraffe}, I., {Chabrier}, G., \& {Gallardo}, J. 2009, \apjl, 702, L27

\bibitem[{{Basri} {et~al.}(2005){Basri}, {Borucki}, \& {Koch}}]{basri05}
{Basri}, G., {Borucki}, W.~J., \& {Koch}, D. 2005, New Astronomy Review, 49,
  478

\bibitem[{{Baumann} {et~al.}(2010){Baumann}, {Ram\'irez}, {Mel\'endez},
  {Asplund}, \& {Lind}}]{baumann10}
{Baumann}, P., {Ram\'irez}, I., {Mel\'endez}, J., {Asplund}, M., \& {Lind}, K.
  2010, {A}\&A, in press

\bibitem[{{Bensby} \& {Feltzing}(2010)}]{bensby10:rio}
{Bensby}, T. \& {Feltzing}, S. 2010, in IAU Symposium, Vol. 265, IAU Symposium,
  ed. {K.~Cunha, M.~Spite, \& B.~Barbuy}, 300--303

\bibitem[{{Bensby} {et~al.}(2003){Bensby}, {Feltzing}, \&
  {Lundstr{\"o}m}}]{bensby03}
{Bensby}, T., {Feltzing}, S., \& {Lundstr{\"o}m}, I. 2003, \aap, 410, 527

\bibitem[{{Bensby} {et~al.}(2005){Bensby}, {Feltzing}, {Lundstr{\"o}m}, \&
  {Ilyin}}]{bensby05}
{Bensby}, T., {Feltzing}, S., {Lundstr{\"o}m}, I., \& {Ilyin}, I. 2005, \aap,
  433, 185

\bibitem[{{Bensby} {et~al.}(2011){Bensby}, {Feltzing}, \& {Oey}}]{bensby11}
{Bensby}, T., {Feltzing}, S., \& {Oey}, M.~S. 2011, in preparation

\bibitem[{{Bensby} {et~al.}(2007{\natexlab{a}}){Bensby}, {Oey}, {Feltzing}, \&
  {Gustafsson}}]{bensby07:hercules}
{Bensby}, T., {Oey}, M.~S., {Feltzing}, S., \& {Gustafsson}, B.
  2007{\natexlab{a}}, \apjl, 655, L89

\bibitem[{{Bensby} {et~al.}(2007{\natexlab{b}}){Bensby}, {Zenn}, {Oey}, \&
  {Feltzing}}]{bensby07}
{Bensby}, T., {Zenn}, A.~R., {Oey}, M.~S., \& {Feltzing}, S.
  2007{\natexlab{b}}, \apjl, 663, L13

\bibitem[{{Casagrande} {et~al.}(2010){Casagrande}, {Ram{\'{\i}}rez},
  {Mel{\'e}ndez}, {Bessell}, \& {Asplund}}]{casagrande10}
{Casagrande}, L., {Ram{\'{\i}}rez}, I., {Mel{\'e}ndez}, J., {Bessell}, M., \&
  {Asplund}, M. 2010, \aap, 512, 54

\bibitem[{{Chen} \& {Zhao}(2006)}]{chen06}
{Chen}, Y.~Q. \& {Zhao}, G. 2006, \aj, 131, 1816

\bibitem[{{Ciesla}(2008)}]{ciesla08}
{Ciesla}, F.~J. 2008, Meteoritics and Planetary Science, 43, 639

\bibitem[{{Clegg} {et~al.}(1981){Clegg}, {Tomkin}, \& {Lambert}}]{clegg81}
{Clegg}, R.~E.~S., {Tomkin}, J., \& {Lambert}, D.~L. 1981, \apj, 250, 262

\bibitem[{{D'Antona} \& {Mazzitelli}(1994)}]{dantona94}
{D'Antona}, F. \& {Mazzitelli}, I. 1994, \apjs, 90, 467

\bibitem[{{Ecuvillon} {et~al.}(2006){Ecuvillon}, {Israelian}, {Santos},
  {Mayor}, \& {Gilli}}]{ecuvillon06}
{Ecuvillon}, A., {Israelian}, G., {Santos}, N.~C., {Mayor}, M., \& {Gilli}, G.
  2006, \aap, 449, 809

\bibitem[{{Edvardsson} {et~al.}(1993){Edvardsson}, {Andersen}, {Gustafsson},
  {Lambert}, {Nissen}, \& {Tomkin}}]{edvardsson93}
{Edvardsson}, B., {Andersen}, J., {Gustafsson}, B., {et~al.} 1993, \aap, 275,
  101

\bibitem[{{Fabbian} {et~al.}(2009){Fabbian}, {Asplund}, {Barklem}, {Carlsson},
  \& {Kiselman}}]{fabbian09}
{Fabbian}, D., {Asplund}, M., {Barklem}, P.~S., {Carlsson}, M., \& {Kiselman},
  D. 2009, \aap, 500, 1221

\bibitem[{{Fedele} {et~al.}(2010){Fedele}, {van den Ancker}, {Henning},
  {Jayawardhana}, \& {Oliveira}}]{fedele10}
{Fedele}, D., {van den Ancker}, M.~E., {Henning}, T., {Jayawardhana}, R., \&
  {Oliveira}, J.~M. 2010, \aap, 510, 72

\bibitem[{{Feldman}(1992)}]{feldman92}
{Feldman}, U. 1992, \physscr, 46, 202

\bibitem[{{Feltzing} \& {Bensby}(2008)}]{feltzing08}
{Feltzing}, S. \& {Bensby}, T. 2008, Physica Scripta Volume T, 133, 014031

\bibitem[{{Fischer} \& {Valenti}(2005)}]{fischer05}
{Fischer}, D.~A. \& {Valenti}, J. 2005, \apj, 622, 1102

\bibitem[{{Gonzalez}(1997)}]{gonzalez97}
{Gonzalez}, G. 1997, \mnras, 285, 403

\bibitem[{{Gonzalez}(1998)}]{gonzalez98}
{Gonzalez}, G. 1998, \aap, 334, 221

\bibitem[{{Gonzalez}(2008)}]{gonzalez08}
{Gonzalez}, G. 2008, \mnras, 386, 928

\bibitem[{{Gonzalez} {et~al.}(2010){Gonzalez}, {Carlson}, \&
  {Tobin}}]{gonzalez10}
{Gonzalez}, G., {Carlson}, M., \& {Tobin}, R.~W. 2010, MNRAS, in press

\bibitem[{{Gonzalez} {et~al.}(2001){Gonzalez}, {Laws}, {Tyagi}, \&
  {Reddy}}]{gonzalez01}
{Gonzalez}, G., {Laws}, C., {Tyagi}, S., \& {Reddy}, B.~E. 2001, \aj, 121, 432

\bibitem[{{Gonzalez Hernandez} {et~al.}(2010){Gonzalez Hernandez}, {Israelian},
  {Santos}, {Sousa}, {Delgado-Mena}, {Neves}, \& {Udry}}]{gonzalez-hernandez10}
{Gonzalez Hernandez}, J.~I., {Israelian}, G., {Santos}, N.~C., {et~al.} 2010,
  ApJ, in press

\bibitem[{{Gustafsson} {et~al.}(2010){Gustafsson}, {Mel{\'e}ndez}, {Asplund},
  \& {Yong}}]{gustafsson10}
{Gustafsson}, B., {Mel{\'e}ndez}, J., {Asplund}, M., \& {Yong}, D. 2010, \apss,
  328, 185

\bibitem[{{Heiter} \& {Luck}(2003)}]{heiter03}
{Heiter}, U. \& {Luck}, R.~E. 2003, \aj, 126, 2015

\bibitem[{{Hekker} \& {Mel{\'e}ndez}(2007)}]{hekker07}
{Hekker}, S. \& {Mel{\'e}ndez}, J. 2007, \aap, 475, 1003

\bibitem[{{H{\'e}noux}(1998)}]{henoux98}
{H{\'e}noux}, J. 1998, Space Science Reviews, 85, 215

\bibitem[{{Israelian} {et~al.}(2009){Israelian}, {Delgado Mena}, {Santos},
  {Sousa}, {Mayor}, {Udry}, {Dom{\'{\i}}nguez Cerde{\~n}a}, {Rebolo}, \&
  {Randich}}]{israelian09}
{Israelian}, G., {Delgado Mena}, E., {Santos}, N.~C., {et~al.} 2009, \nat, 462,
  189

\bibitem[{{Lodders}(2003)}]{lodders03}
{Lodders}, K. 2003, \apj, 591, 1220

\bibitem[{{Luck} \& {Heiter}(2006)}]{luck06}
{Luck}, R.~E. \& {Heiter}, U. 2006, \aj, 131, 3069

\bibitem[{{Mamajek}(2009)}]{mamajek09}
{Mamajek}, E.~E. 2009, in American Institute of Physics Conference Series, Vol.
  1158, American Institute of Physics Conference Series, ed. {T.~Usuda,
  M.~Tamura, \& M.~Ishii}, 3--10

\bibitem[{{Marcy} {et~al.}(2005){Marcy}, {Butler}, {Fischer}, {Vogt}, {Wright},
  {Tinney}, \& {Jones}}]{marcy05}
{Marcy}, G., {Butler}, R.~P., {Fischer}, D., {et~al.} 2005, Progress of
  Theoretical Physics Supplement, 158, 24

\bibitem[{{Mel{\'e}ndez} {et~al.}(2009){Mel{\'e}ndez}, {Asplund}, {Gustafsson},
  \& {Yong}}]{melendez09:twins}
{Mel{\'e}ndez}, J., {Asplund}, M., {Gustafsson}, B., \& {Yong}, D. 2009, \apjl,
  704, L66

\bibitem[{{Mel{\'e}ndez} {et~al.}(2010){Mel{\'e}ndez}, {Ram{\'{\i}}rez},
  {Casagrande}, {Asplund}, {Gustafsson}, {Yong}, {Do Nascimento}, {Castro}, \&
  {Bazot}}]{melendez09:lithium}
{Mel{\'e}ndez}, J., {Ram{\'{\i}}rez}, I., {Casagrande}, L., {et~al.} 2010,
  \apss, 328, 193

\bibitem[{{Meyer}(2009)}]{meyer09}
{Meyer}, M.~R. 2009, in IAU Symposium, Vol. 258, IAU Symposium, ed. E.~E.
  {Mamajek}, D.~R. {Soderblom}, \& R.~F.~G. {Wyse}, 111--122

\bibitem[{{Neves} {et~al.}(2009){Neves}, {Santos}, {Sousa}, {Correia}, \&
  {Israelian}}]{neves09}
{Neves}, V., {Santos}, N.~C., {Sousa}, S.~G., {Correia}, A.~C.~M., \&
  {Israelian}, G. 2009, \aap, 497, 563

\bibitem[{{Nordlund}(2009)}]{nordlund09}
{Nordlund}, {\AA.}. 2009, ApJ, submitted (arXiv:astro-ph/0908.3479)

\bibitem[{{Pasquini} {et~al.}(2007){Pasquini}, {D{\"o}llinger}, {Weiss},
  {Girardi}, {Chavero}, {Hatzes}, {da Silva}, \& {Setiawan}}]{pasquini07}
{Pasquini}, L., {D{\"o}llinger}, M.~P., {Weiss}, A., {et~al.} 2007, \aap, 473,
  979

\bibitem[{{Pinsonneault} {et~al.}(2001){Pinsonneault}, {DePoy}, \&
  {Coffee}}]{pinsonneault01}
{Pinsonneault}, M.~H., {DePoy}, D.~L., \& {Coffee}, M. 2001, \apjl, 556, L59

\bibitem[{{Ram{\'{\i}}rez} {et~al.}(2007){Ram{\'{\i}}rez}, {Allende~Prieto}, \&
  {Lambert}}]{ramirez07}
{Ram{\'{\i}}rez}, I., {Allende~Prieto}, C., \& {Lambert}, D.~L. 2007, \aap,
  465, 271

\bibitem[{{Ram{\'{\i}}rez} {et~al.}(2009){Ram{\'{\i}}rez}, {Mel{\'e}ndez}, \&
  {Asplund}}]{ramirez09}
{Ram{\'{\i}}rez}, I., {Mel{\'e}ndez}, J., \& {Asplund}, M. 2009, \aap, 508, L17

\bibitem[{{Reddy} {et~al.}(2006){Reddy}, {Lambert}, \& {Allende
  Prieto}}]{reddy06}
{Reddy}, B.~E., {Lambert}, D.~L., \& {Allende Prieto}, C. 2006, \mnras, 367,
  1329

\bibitem[{{Reddy} {et~al.}(2003){Reddy}, {Tomkin}, {Lambert}, \&
  {Allende~Prieto}}]{reddy03}
{Reddy}, B.~E., {Tomkin}, J., {Lambert}, D.~L., \& {Allende~Prieto}, C. 2003,
  \mnras, 340, 304

\bibitem[{{Ryan}(2000)}]{ryan00}
{Ryan}, S.~G. 2000, \mnras, 316, L35

\bibitem[{{Sadakane} {et~al.}(2002){Sadakane}, {Ohkubo}, {Takeda}, {Sato},
  {Kambe}, \& {Aoki}}]{sadakane02}
{Sadakane}, K., {Ohkubo}, M., {Takeda}, Y., {et~al.} 2002, \pasj, 54, 911

\bibitem[{{Santos} {et~al.}(2004){Santos}, {Israelian}, \& {Mayor}}]{santos04}
{Santos}, N.~C., {Israelian}, G., \& {Mayor}, M. 2004, \aap, 415, 1153

\bibitem[{{Sousa} {et~al.}(2008){Sousa}, {Santos}, {Mayor}, {Udry},
  {Casagrande}, {Israelian}, {Pepe}, {Queloz}, \& {Monteiro}}]{sousa08}
{Sousa}, S.~G., {Santos}, N.~C., {Mayor}, M., {et~al.} 2008, \aap, 487, 373

\bibitem[{{Takeda}(2007)}]{takeda07:abundances}
{Takeda}, Y. 2007, \pasj, 59, 335

\bibitem[{{Takeda} {et~al.}(2008){Takeda}, {Sato}, \& {Murata}}]{takeda08}
{Takeda}, Y., {Sato}, B., \& {Murata}, D. 2008, \pasj, 60, 781

\bibitem[{{Udry} \& {Santos}(2007)}]{udry07}
{Udry}, S. \& {Santos}, N.~C. 2007, \araa, 45, 397

\bibitem[{{Wuchterl} \& {Tscharnuter}(2003)}]{wuchterl03}
{Wuchterl}, G. \& {Tscharnuter}, W.~M. 2003, \aap, 398, 1081

\end{thebibliography}

\end{document}